\documentclass[11pt,twoside,onecolumn]{article}
\usepackage[]{latexsym}
\usepackage{epsfig}
\usepackage{amsmath,amssymb}
\newcommand{\be}{\begin{eqnarray}}
\newcommand{\ee}{\end{eqnarray}}

\newcommand{\lp}{\ell_{\rm P}}
\newcommand{\mpl}{M_{\rm P}}
\renewcommand{\lg}{\ell_{\rm G}}
\newcommand{\mg}{M_{\rm G}}
\newcommand{\rh}{R_{\rm H}}

\newcommand{\mc}{M_{\rm C}}

\title{\bf Brane-world stars from minimal geometric deformation, and black holes}
\author{Roberto~Casadio$^{a,b}$\thanks{casadio@bo.infn.it}
$\,$ and
Jorge~Ovalle$^{c}$\thanks{jovalle@usb.ve}
\\
\null
\\
$^a${\em Dipartimento di Fisica e Astronomia, Universit\`a di Bologna}
\\
{\em via Irnerio~46, 40126 Bologna, Italy}
\\
\\
$^b${\em Istituto Nazionale di Fisica Nucleare, Sezione di Bologna}
\\
{\em via B.~Pichat~6/2, 40127 Bologna, Italy}
\\
$^c${\em Departamento de Fisica, Universidad Sim\'on Bol\'ivar}
\\
{\em Apartado 89000, Caracas 1080A, Venezuela}
}
\begin{document}
\maketitle
\begin{abstract}
We build analytical models of spherically symmetric stars in the brane-world,
in which the external space-time contains both an ADM mass and a tidal charge.
In order to determine the interior geometry, we apply the principle of minimal
geometric deformation, which allows one to map General Relativistic solutions
to solutions of the effective four-dimensional brane-world equations.
We further restrict our analysis to stars with a radius linearly related to the
total General Relativistic mass, and obtain a general relation between the
latter, the brane-world ADM mass and the tidal charge.
In these models, the value of the star's radius can then be taken to zero smoothly,
thus obtaining brane-world black hole metrics with a tidal charge
solely determined by the mass of the source and the brane tension.
General conclusions regarding the minimum mass for semiclassical black holes
will also be drawn.
\end{abstract}
%
%
%
%
%
%
%
\section{Introduction}
\label{intro}
It is well-known that black holes (BHs) are unstable in four (and higher)
dimensions because they emit Hawking particles~\cite{hawking},
a quantum effect signalled by the trace anomaly of the radiation field
on the classical vacuum black hole background~\cite{CF,birrell,holo}.
A complete description of BH evaporation would indeed require solving
the semiclassical Einstein equations with the back-reaction of the evaporation
flux on the metric, an extremely hard task.
In the context of the Randall-Sundrum (RS) Brane-World (BW)
models~\cite{RS}, it was shown in Refs.~\cite{BGM,kofinas}
that the collapse of a homogeneous star leads to a non-static exterior,
contrary to what happens in four-dimensional General Relativity (GR),
and a possible exterior was later found which is radiative~\cite{dad}.
If one regards BHs as the natural end-states of the collapse,
one may conclude that classical BHs in the BW should
suffer of the same problem as BHs in GR:
no static configuration for their exterior might be allowed~\cite{tanaka1,fabbri}.
Nonetheless, a static configuration was recently found in Ref.~\cite{wiseman}
(see also Refs.~\cite{dejan,page} for more explicit examples of static metrics),
thus keeping the debate open as to what we should consider a realistic
physical description of stars and BHs in models with extra spatial dimensions.
\par
Solving the full five-dimensional Einstein field equations with appropriate
sources (BH plus brane) is indeed a formidable task (see, e.g.~\cite{cmazza,darocha},
and References therein), and finding analytically manageable solutions
does not seem likely.
In fact, the recently found five-dimensional BH metrics~\cite{wiseman,dejan,page}
are only partly known in their analytic form, and no uniqueness theorem has
been proven to ensure different solutions do not exist.
Moreover, a complete analysis of the possible corresponding matter sources
is still missing, which makes them just a firsts step for addressing the topic of
stars in the BW.
The two issues are in fact related. 
One typically builds a star model by choosing the star interior, which should
solve (five-dimensional) Einstein equations as well,
and then matching it with the exterior vacuum geometry,
a procedure which definitely benefits of a fully analytical description
of all quantities involved.
Without a uniqueness theorem (a BW {\em no-hair\/} theorem), it is still possible
that different interiors generate exterior metrics that differs from those in
Refs.~\cite{wiseman,dejan,page}.
Moreover, even if we worked with an exact five-dimensional solution,
we might still obtain different geometries in our observed universe,
since there are many ways to embed a four-dimensional brane in the
five-dimensional bulk.
It seems therefore safe to conclude that the space-time geometry of BW stars
remains an open problem, and that finding both four- and five-dimensional
analytical solutions, also employing alternative approaches to the one in
Refs.~\cite{wiseman,dejan,page}, should help to understand their physics
and work toward the formulation of a uniqueness theorem.
It is also not clear to what extent the results in Refs.~\cite{wiseman,dejan,page}
can be applied to small BHs, that is BHs with a mass near the fundamental
scale of gravity, for which quantum (or semiclassical) corrections could be large
enough to deform the geometry significantly.
This is another crucial aspect in light of possible microscopic BH phenomenology
at the LHC~\cite{cavaglia}.
\par
The approach we will follow here is to start from a consistent
description of BW stars, with a static radius larger than the would-be horizon
radius of the system, and an outer geometry containing a tidal
charge, which is usually regarded as independent from the ADM mass~\cite{dadhich}.
We will then employ a (mathematical) limit, which (formally~\footnote{Note the
the ADM mass~\eqref{gMadm} and charge~\eqref{gqM} do not explicitly
depend on the star radius, so that the latter can be set to any value.})
takes the star's radius to zero, and allows us to recover a space-time with a
BH event horizon.
Of course, there is no guarantee that such a mathematical process will
reproduce the physics of a real gravitational collapse, or any sort of
BH formation.
We shall nonetheless find that the tidal charge is henceforth naturally related
to the ADM mass, and this will have interesting consequences which may help
to better understand BH physics in the BW.
\subsection{BW basics}
We recall that, in the RS model, our Universe is a co-dimension one
four-dimensional hypersurface of vacuum energy density
$\rho_{\rm RS}=\mpl\,\sigma/\lp$
(related to the bulk cosmological constant and warp factor).
We shall usually display the Newton constant $G_{\rm N}=\lp/\mpl$ explicitly
and denote by $\lg\gg\lp$ and $\mg\ll\mpl$
the fundamental five-dimensional length and mass ($\hbar=\mpl\,\lp=\mg\,\lg$).
The ``brane density'' parameter $\sigma$ thus has dimensions of an inverse
squared length, namely $\sigma\simeq\lg^{-2}$~\cite{RS}.
\par
In Gaussian normal coordinates $x^A=(x^\mu,y)$, where $y$ is the
extra-dimensional coordinate with the brane located at $y=0$
(capitol letters run from $0$~to~$4$ and Greek letters from $0$~to~$3$),
the five-dimensional metric can then be expanded near the brane
as~\cite{maart}
\be
g^5_{AB}
\simeq
\left.g^5_{AB}\right|_{y=0}
+\left.2\,K_{AB}\right|_{y=0}\, y
+\left.\pounds_{\hat n}K_{AB}\right|_{y=0}\,y^2
\ ,
\label{g5}
\ee
where $K_{AB}$ is the extrinsic curvature of the brane,
and $\pounds_{\hat n}$ the Lie derivative along the unitary four-vector
$\hat n$ orthogonal to the brane.
Junction conditions at the brane lead to~\cite{shiromizu}
\be
K_{\mu\nu}\sim
T_{\mu\nu} -\frac{1}{3}\,\left(T-\sigma\right)\,g_{\mu\nu}
\ ,
\label{K}
\ee
where $T_{\mu\nu}$ is the stress tensor of the matter localized
on the brane, and~\cite{maart}
\be
\pounds_{\hat n} K_{\mu\nu}\sim {\cal E}_{\mu\nu}+f(T)_{\mu\nu}
\ ,
\label{f}
\ee
where ${\cal E}_{\mu\nu}$ is the projection of the Weyl tensor on the
brane and $f(T)_{\mu\nu}$ a tensor which depends on $T_{\mu\nu}$ and
$\sigma$.
\subsection{Junction conditions}
The well-known junction conditions in GR~\cite{israel} allow for
(step-like) discontinuities in the stress tensor $T_{\mu\nu}$
(for example, across the surface of a star),
keeping the first and second fundamental forms continuous.
For thin (Dirac $\delta$-like) surfaces, a step-like discontinuity of the
extrinsic curvature $K_{\mu\nu}$ orthogonal to the surface is also allowed if
the metric remains continuous~\cite{israel}.
\par
However, since a brane in RS is itself a thin surface,
it generates an orthogonal discontinuity of the extrinsic curvature
$K_{AB}$ in five dimensions.
Moreover, discontinuities in the localised matter stress tensor $T_{\mu\nu}$
would induce discontinuities in the extrinsic curvature~\eqref{K}
tangential to the brane, which should appear in the
five-dimensional metric~\eqref{g5}.
Such discontinuities of the metric $g_{AB}$ are not allowed by the regularity of
five-dimensional geodesics.
Moreover, because of the second order term in Eq.~\eqref{g5}, and considering
Eq.~\eqref{f}, we can not allow the projected Weyl tensor to be discontinuous
on the brane either.
One can understand the physical meaning of the above regularity requirement
by considering that, in a microscopic description of the BW, matter should be smooth
along the extra dimension, yet localized on the brane (say, within a width
of order $\sigma^{-1/2}$, see e.g.~Ref.~\cite{g2}).
In any such description, the continuity of five-dimensional geodesics must
then hold and, in order to build a physical model of a star, one has
to smooth both the matter stress tensor and the projected Weyl
tensor across the surface of the star along the brane.
\subsection{BW stars and MGD}
In our analysis of BW stars, we shall employ the effective four-dimensional
equations of Ref.~\cite{shiromizu}.
In general, these equations represent an open system on the brane,
because the contribution ${\cal E}_{\mu\nu}$ from the bulk Weyl tensor
remains undetermined after projecting Einsten's equations onto the brane,
and identifying specific solutions requires more information on
the bulk geometry and a better understanding of how our four-dimensional
space-time is embedded in the bulk.
\par
Nonetheless, it is possible to generate the BW version of every 
spherically symmetric GR solution through the {\it minimal geometric deformation\/}
(MGD) approach~\cite{jovalle2009}. 
This method is based on requiring that any BW solution must reduce
to a corresponding four-dimensional GR solution at low energies
(that is, energy densities much smaller than $\sigma$).
In practical terms, from the point of view of a brane observer,
the existence of gravity in the extra dimension induces a geometric
deformation in the radial metric component, which is the source of
anisotropy on the brane.
When a solution of the four-dimensional Einstein equations
is considered as a possible solution of the BW system, 
the geometric deformation produced by extra-dimensional effects
is minimized, and the open system of effective BW equations is
automatically satisfied.
This approach was successfully used to generate physically
acceptable interior solutions for stellar systems~\cite{jovalle07},
and even exact solutions were found in Ref.~\cite{jovalle207}.
\subsection{Outline}
The remaining of the paper is organised as follows:
In Section~\ref{star}, we shall briefly review the general BW formalism 
and description of spherically symmetric BW stars, with particular emphasis
on the principle of MGD, as a means to obtain interior solutions,
and matching conditions at the star's surface.
The exterior metric will naturally be chosen to contain a tidal term and
be of the form given in Ref.~\cite{dadhich}. 
In Section~\ref{ADM_q}, we will obtain several explicit relations
between the ADM mass and tidal charge determined by specific
choices of the star distributions and compactness. 
In Section~\ref{bh}, we shall reconsider such BW metrics
for the star exterior and impose auxiliary conditions in order
to enforce a ``BH limit'', as was first done in Ref.~\cite{covalle}.
We shall show that such conditions relate the outer tidal term
to the ADM mass $\mathcal{M}$ and can be used to obtain
a minimum mass for semiclassical BW BHs.
Finally, we shall briefly discuss our findings in Section~\ref{conc}.
\section{BW star models}
\label{star}
The simplest GR model of a star collapsing to form a BH,
the Oppenheimer-Snyder (OS) model~\cite{OS},
is known to be unreliable in the RS scenario~\cite{BGM}.
In Ref.~\cite{gercas}, modifications to the OS model based on the
Tolman geometry~\cite{tolman} were therefore studied,
showing that it converges to the GR solution at low energies
and both the matter stress tensor $T_{\mu\nu}$ and the projected
Weyl tensor ${\cal E}_{\mu\nu}$ are continuous, as discussed above.
It was then shown that, for a collapsing star of small energy density
on an asymptotically flat brane, the full five-dimensional dynamics
of the innermost region determines the entire evolution of the BW
system.
In particular,  the total energy of the system is conserved~\footnote{We
note in passing this is the main assumption in the microcanonical
treatment of the black hole evaporation (see, e.g.~Refs.~\cite{micro}).}
and the collapsing star ``evaporates'' until the core
experiences a ``rebound'' in the high energy regime
(when its energy density becomes comparable with $\sigma$), 
after which the whole system ``anti-evaporates''.
This suggested the possibility that a collapsing dust star eventually
goes back to the state which it started from.
\par
Unlike the present work, no effort was made in Ref.~\cite{gercas}
to determine solutions in which the inner pressure of the star can
balance both four-dimensional gravitational attraction and
tidal effects to generate a static solution.
In this Section we will study models of stable BW stars,
with density and pressure, and obtain the external metric
by imposing the usual junction conditions at the star's surface
and the principle of MGD.
\subsection{General framework}
In the context of the RS model, five-dimensional gravity induces
modifications in the Einstein's field equations on our $(3+1)$-dimensional
observable universe, the brane, which then take the effective
form~\cite{maart,shiromizu}
\begin{equation}
\label{einst}
G_{\mu\nu}=
-k^2\,T_{\mu\nu}^{T}-\Lambda\, g_{\mu\nu}
\ ,
\end{equation}
where $k^2=8\,\pi\,G_{\rm N}$, $\Lambda$ is the cosmological constant
on the brane (not to be confused with the necessarily non-vanishing
bulk cosmological constant)
and $T_{\mu\nu}^{T}$ the effective energy-momentum tensor
given by 
\begin{equation}
\label{tot}
T_{\mu\nu}^{\;\;T}
=T_{\mu\nu}+\frac{6}{\sigma}\,S_{\mu\nu}
+\frac{1}{8\,\pi}\,{\cal E}_{\mu\nu}
+\frac{4}{\sigma}\,{\cal F}_{\mu\nu}
\ .
\end{equation}
In the above, $T_{\mu\nu}$ is again the usual stress tensor of matter localized
on the brane, $\sigma$ the brane tension described in the Introduction,
$S_{\mu\nu}$ and $\cal{E}_{\mu\nu}$ the high-energy
and non-local (from the point of view of a brane observer)
corrections respectively, and ${\cal F}_{\mu\nu}$ a term which depends
on all stresses in the bulk but the bulk cosmological constant.
\par 
In this work, we shall consider only a cosmological constant in the bulk
(which, by fine tunning, will produce $\Lambda=0$ on the brane), hence 
\be
{\cal F}_{\mu\nu}=0
\ ,
\ee
which implies the conservation equation
\be
\nabla^\nu\,T_{\mu\nu}=0
\ ,
\ee
so that there will be no exchange of energy between the bulk and the brane.
We shall only consider BW matter described by a perfect fluid with four-velocity
$u^\mu$,
and $h_{\mu\nu}=g_{\mu\nu}-u_\mu u_\nu$  will then denote the projection tensor
orthogonal to the fluid lines.
The extra terms $S_{\mu\nu}$ and $\cal{E}_{\mu\nu}$ represent the high-energy
and Kaluza-Klein corrections, respectively, and are given by
\be
\label{s}
S_{\mu\nu}=
\frac{T\,T_{\mu\nu}}{12}
-\frac{T_{\mu\alpha}\,T^\alpha_{\ \nu}}{4}
+\frac{g_{\mu\nu}}{24}
\left[3\,T_{\alpha\beta}\,T^{\alpha\beta}-T^2\right]
\ee
where $T=T_\alpha^{\ \alpha}$, and
\be
\label{e}
k^2\,{\cal E}_{\mu\nu}
=
\frac{6}{\sigma}\left[{\cal U}\left(u_\mu\,u_\nu+\frac{1}{3}\,h_{\mu\nu}\right)
+{\cal P}_{\mu\nu}+{\cal Q}_{(\mu}\,u_{\nu)}\right]
\ ,
&&
\ee
with ${\cal U}$ the bulk Weyl scalar, and ${\cal P}_{\mu\nu}$ and
${\cal Q}_\mu $ the anisotropic stress and energy flux, respectively.
The form of these terms can be further restricted by assuming spherical
symmetry on the brane.
\subsection{Spherically symmetric stars}
\label{ssd}
For our stars, we shall consider spherically symmetric static distributions,
hence $Q_\mu =0$ and
\begin{equation}
{\cal P}_{\mu\nu}
={\cal P}\left(r_\mu\, r_\nu+\frac{1}{3}\,h_{\mu\nu}\right)
\ ,
\end{equation}
where $r_\mu$ is a unit radial vector.
The line element is therefore given, in Schwarzschild-like coordinates,
by
\be
\label{metric}
ds^2
=
e^\nu\, dt^2-e^\lambda\, dr^2
-r^2\left( d\theta^2+\sin^2\theta\,d\phi ^2\right)
\ ,
\ee
where $\nu=\nu(r)$ and $\lambda=\lambda(r)$ are functions of
the areal radius $r$, which ranges from $r=0$ (the star's centre)
to $r=R$ (the star's surface).
\par
The metric~\eqref{metric} must satisfy the effective Einstein field
equations~\eqref{einst}, which, for $\Lambda=0$, explicitly read
\be
\label{ec1}
&&
k^2
\left[ \rho
+\strut\displaystyle\frac{1}{\sigma}\left(\frac{\rho^2}{2}+\frac{6}{k^4}\,\cal{U}\right)
\right]
=
\strut\displaystyle\frac 1{r^2}
-e^{-\lambda }\left( \frac1{r^2}-\frac{\lambda'}r\right)
\\
&&
\label{ec2}
k^2
\strut\displaystyle
\left[p+\frac{1}{\sigma}\left(\frac{\rho^2}{2}+\rho\, p
+\frac{2}{k^4}\,\cal{U}\right)
+\frac{4}{k^4}\frac{\cal{P}}{\sigma}\right]
=
-\frac 1{r^2}+e^{-\lambda }\left( \frac 1{r^2}+\frac{\nu'}r\right)
\\
&&
\label{ec3}
k^2
\strut\displaystyle\left[p
+\frac{1}{\sigma}\left(\frac{\rho^2}{2}+\rho\, p
+\frac{2}{k^4}\cal{U}\right)
-\frac{2}{k^4}\frac{\cal{P}}{\sigma}\right]
=
\frac 14e^{-\lambda }\left[ 2\,\nu''+\nu'^2-\lambda'\,\nu'
+2\,\frac{\nu'-\lambda'}r\right]
\ .
\ee
Moreover,
\be
\label{con1}
p'=-\strut\displaystyle\frac{\nu'}{2}(\rho+p)
\ ,
\ee
where $f'\equiv \partial_r f$.
We then note that four-dimensional GR equations are formally
recovered for $\sigma^{-1}\to 0$, and the conservation equation~\eqref{con1}
then becomes a linear combination of Eqs.~\eqref{ec1}-\eqref{ec3}.
\par
The Israel-Darmois matching conditions~\cite{israel} at the stellar surface
$\Sigma$ of radius $r=R$ give
\be
\label{matching1}
\left[G_{\mu\nu}\,r^\nu\right]_{\Sigma}=0
\ ,
\ee
where $[f]_{\Sigma}\equiv f(r\to R^+)-f(r\to R^-)$.
Using Eq.~\eqref{matching1} and the general field equations~\eqref{einst},
we find
\be
\label{matching2}
\left[T^{T}_{\mu\nu}\,r^\nu\right]_{\Sigma}=0
\ ,
\ee
which, for the specific form of our metric and fluid source, leads to
\be
\label{matching3}
\left[
p+\frac{1}{\sigma}\left(\frac{\rho^2}{2}+\rho\, p
+\frac{2}{k^4}\,\cal{U}\right)+\frac{4}{k^4}\,\frac{\cal{P}}{\sigma}
\right]_{\Sigma}=0
\ .
\ee
Since we assume the star is surrounded by empty space,
$p=\rho=0$ for $r>R$, this matching condition
takes the final form
\be
\label{matchingf}
p_R+\frac{1}{\sigma}\left(\frac{\rho_R^2}{2}+\rho_R\, p_R
+\frac{2}{k^4}\,{\cal U}_R^-\right)
+\frac{4}{k^4}\frac{{\cal P}_R^-}{\sigma}
=
\frac{2}{k^4}\frac{{\cal U}_R^+}{\sigma}+\frac{4}{k^4}\frac{{\cal P}_R^+}{\sigma}
\ ,
\ee
where $f_R^\pm\equiv f(r\to R^\pm)$, with $p_R\equiv p_R^-$
and $\rho_R\equiv \rho_R^-$.
\par
Eq.~\eqref{matchingf} gives the general matching condition
for any static spherical BW star~\cite{gm,gergely2006}.
In the limit $\sigma^{-1}\rightarrow 0$, we obtain the well-known GR
matching condition $p_R =0$ at the star surface.
In the particular case of the Schwarzschild exterior,
${\cal U}^+={\cal P}^+ =0$, the matching condition~\eqref{matchingf}
becomes
\be
\label{matchingfS}
p_R+\frac{1}{\sigma}\left(\frac{\rho_R^2}{2}+\rho_R\, p_R
+\frac{2}{k^4}\,{\cal U}_R^-\right)
+\frac{4}{k^4}\frac{{\cal P}_R^-}{\sigma} = 0
\ .
\ee
This clearly shows that, because of the presence of
${\cal U}_R^-$ and ${\cal P}_R^-$, the matching conditions
do not have a unique solution in the BW.
\subsection{MGD stars}
\begin{figure}[h]
\centering
\hspace{-3.5cm}
\epsfxsize=7cm
\epsfbox{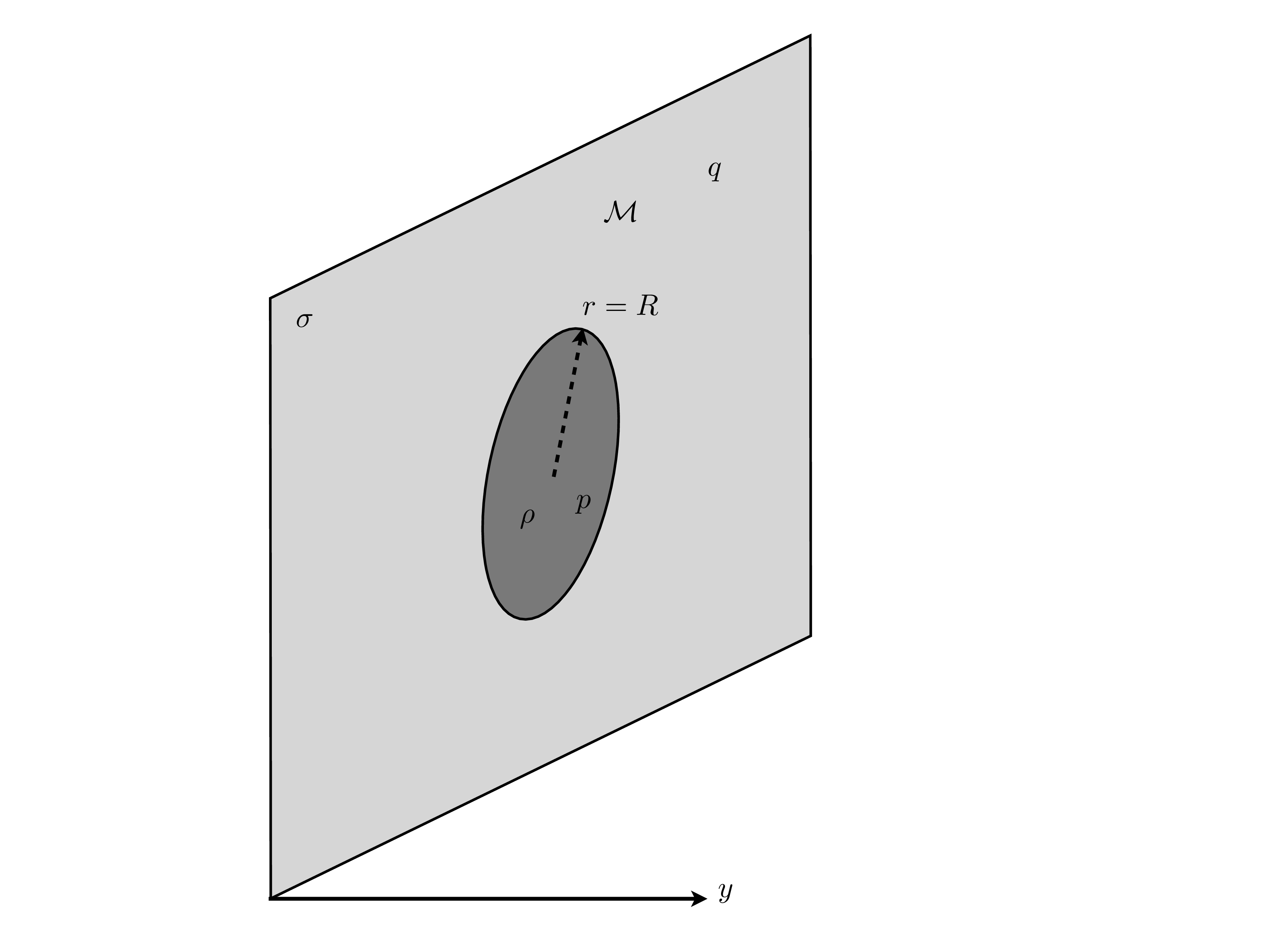}
\caption{Schematic picture of a BW star: the interior is characterized by
the density $\rho$, pressure $p$ and star radius $R$, the exterior geometry
by the ADM mass ${\cal M}$ and tidal charge $q$.
The brane has tension $\sigma$, fine-tuned with the bulk cosmological
constant, so that the brane cosmological constant $\Lambda=0$. 
\label{BWstar}}
\end{figure}
We now proceed to describe in general the model of a BW star starting
from its interior, which will be obtained from known GR solutions
by means of the MGD approach.
We then apply the matching conditions at the star
surface with a prescribed external geometry.
\par
A pictorial representation of the system and relevant variables
can be seen in Fig.~\ref{BWstar}.
\subsubsection{General star interior}
It is easy to see that the field equations~\eqref{ec1}-\eqref{ec3}
can be rewritten as
\be
\label{usual}
&&
e^{-\lambda}
=
1-\frac{k^2}{r}\int_0^r
x^2
\left[
\rho+\frac{1}{\sigma}\left(\frac{\rho^2}{2}+\frac{6}{k^4}\,\cal{U}\right)
\right]
dx
\\
\label{pp}
&&
\frac{1}{k^2}\,\frac{{\cal P}}{\sigma}
=
\frac{1}{6}\left(G_{\ 1}^{1}-G_{\ 2}^2\right)
\\
\label{uu}
&&
\frac{6}{k^4}\,\frac{{\cal U}}{\sigma}
=
-\frac{3}{\sigma}\left(\frac{\rho^2}{2}+\rho\,p\right)
+\frac{1}{k^2}\left(2\,G_{\ 2}^2+G_{\ 1}^1\right)-3\,p
\ ,
\ee
with
\be
\label{g11}
G_{\ 1}^1
=
-\frac 1{r^2}+e^{-\lambda }\left( \frac 1{r^2}+\frac{\nu'}r\right)
\ee
and
\be
\label{g22}
G_{\ 2}^2
=
\frac 14\,e^{-\lambda }\left( 2\,\nu''+\nu'^2-\lambda'\,\nu'+2 \frac{\nu'-\lambda'}r
\right)
\ .
\ee 
\par
Eq.~\eqref{usual} is an integro-differential equation for the function
$\lambda=\lambda(r)$, something completely different from the GR case,
and a direct consequence of the non-locality of the BW equations.
The only general solution known for this equation is given by~\cite{jovalle2009}
\be
\label{edlrwss}
e^{-\lambda}
&\!\!=\!\!&
{1-\frac{k^2}{r}\int_0^r
x^2\,\rho\,dx}
+\underbrace{e^{-I}\int_0^r\frac{e^I}{\frac{\nu'}{2}+\frac{2}{x}}
\left[H(p,\rho,\nu)+\frac{k^2}{\sigma}\left(\rho^2+3\,\rho \,p\right)\right]
dx}_{\rm Geometric\ deformation}
\nonumber
\\
&\!\!\equiv\!\!&
\mu(r)+f(r)
\ ,
\ee
where
\be
\label{finalsol}
H(p,\rho,\nu)
\equiv
3\,k^2\,p
-\left[\mu'\left(\frac{\nu'}{2}+\frac{1}{r}\right)
+\mu\left(\nu''+\frac{\nu'^2}{2}+\frac{2\nu'}{r}+\frac{1}{r^2}\right)
-\frac{1}{r^2}\right]
\ ,
\ee
and
\be
\label{I}
I
\equiv
\int\frac{\left(\nu''+\frac{{\nu'}^2}{2}+\frac{2\nu'}{r}+\frac{2}{r^2}\right)}
{\left(\frac{\nu'}{2}+\frac{2}{r}\right)}\,dr
\ .
\ee
Note that the function 
\be
\label{standardGR}
\mu(r)
\equiv
1-\frac{k^2}{r}\int_0^r x^2\,\rho\, dx
=1-\frac{2\,m(r)}{r}
\ee
contains the usual GR mass function $m=\lp\,M/\mpl$,
whereas the function $H(p,\rho,\nu)$ encodes the anisotropic effects on $p$,
$\rho$ and $\nu$ due to bulk gravity.
\par
A crucial observation is now that, when a given GR solution is considered
as a candidate solution for the BW system of Eqs.~\eqref{ec1}-\eqref{con1}
[or, equivalently, Eq.~\eqref{con1} along with Eqs.~\eqref{usual}-\eqref{uu}],
one obtains
\be
H(p,\rho,\nu)=0
\ .
\ee
Consequently, the geometric deformation undergone by the radial metric
component, explicitly shown as $f=f(r)$ in Eq.~(\ref{edlrwss}), is minimal.
This provides the foundation for the MGD approach~\cite{jovalle2009}.
In particular, this approach yields a minimal deformation $f=f^*(r)$
given by Eq.~\eqref{edlrwss} with $H=0$, that is
\be
\label{fsolutionmin}
f^{*}(r)
=
\frac{2\,k^2}{\sigma}\,
e^{-I(r)}\int_0^r
\frac{x\,e^{I(x)}}{x\,\nu'+4}\left(\rho^2+3\,\rho\, p\right)
dx
\ .
\ee
Note that Eq.~\eqref{fsolutionmin} implies the geometric deformation
$f^{*}(r)\geq0$ for $r>0$, hence it always reduces the effective interior mass
[see also Eqs.~\eqref{reglambda} and \eqref{massfunction} below].
On the other hand, Eq.~(\ref{fsolutionmin}) represents a minimal
deformation in the sense that all sources of the deformation have been removed,
except for those produced by the density and pressure, which will always
be present in a realistic stellar  distribution~\footnote{There is a MGD solution
in the case of a dust cloud, with $p=0$, but we will not consider it in the present work.}.
It is worth emphasising that the geometric deformation $f(r)$ shown in
Eq.~\eqref{edlrwss} indeed ``distorts'' the GR solution given in Eq.~\eqref{standardGR}.
The function $f^{*}(r)$ shown in Eq.~\eqref{fsolutionmin} will therefore produce,
from the GR point of view, a ``minimal distortion'' for any GR solution
one wishes to consider.
\subsubsection{Star exterior with tidal charge}
In four dimensions, it is well known that the Schwarzschild metric represents
the unique exterior solution for a spherically symmetric distribution of matter in GR.
This situation changes in the BW, where extra-dimensional gravity effects imply
that the Schwarzschild metric no longer represents a valid exterior geometry. 
\par
In order to investigate how the MGD approach applied to stellar interiors
affects the matching conditions at the star surface, and the corresponding
exterior geometry, we consider a solution to the effective four-dimensional
vacuum Einstein equations~\cite{shiromizu}, namely a metric such that
\be
R_{\mu\nu}-\frac{1}{2}\,g_{\mu\nu}\,R^\alpha_{\ \alpha}
=
\mathcal{E}_{\mu\nu}
\qquad
\Rightarrow
\qquad
R^\alpha_{\ \alpha}
=0
\ .
\ee
This will represent the geometry outside a spherically symmetric stellar BW source,
where we recall that extra-dimensional effects are contained in the projected Weyl
tensor $\mathcal{E}_{\mu\nu}$.
Only a few analytical solutions are known~\cite{wiseman,dejan,dadhich,page,cfabbri}, 
one of them being the tidally charged metric of Ref.~\cite{dadhich}. 
This metric is of the form in Eq.~\eqref{metric} with
\be
\lambda\equiv\lambda_+=-\nu\equiv-\nu_+
\ ,
\ee
and
\be
e^{\nu_+}=1-\frac{2\,\lp\,{\cal M}}{\mpl\,r}-\frac{q}{r^2}
\ ,
\label{tidalg}
\ee
corresponding to
\be
\label{exteriorWayl}
{\cal U}^+=-\frac{{\cal P}^+}{2}
=\frac{4\,\pi\,\sigma\,q}{3\,r^4}
\ ,
\ee
where $\cal{M}$ is the BW ADM mass and $q$ is the tidal charge.
It represents the simplest extension of the Schwarzschild solution,
and has been extensively studied, e.g.~in \cite{acmo,bcLHC,CH} and References
therein.
\par
It is important to remark that the tidal charge $q$ and ADM mass ${\cal M}$
are usually treated as independent quantities.
However, by studying the interior BW solution of compact stars, 
it will be possible to obtain a relationship between $q$ and ${\cal M}$
from the usual Israel's junction conditions~\cite{israel}.
\subsubsection{Matching conditions}
The interior geometry will be represented by a generic solution of the
form~(\ref{metric}), with the non-trivial metric element in Eq.~\eqref{edlrwss},
\be
\label{reglambda}
e^{-\lambda}=1-\frac{2\,\tilde{m}(r)}{r}
\ ,
\ee
where the interior mass function $\tilde{m}$ is given by the MGD
prescription~\eqref{fsolutionmin},
\be
\label{massfunction}
\tilde{m}(r)
=
m(r)-\frac{r}{2}\,f^{*}(r), 
\ee
and $m=m(r)$ is the usual GR mass function, as shown in Eq.~\eqref{standardGR}.
We then consider the junction conditions at the star surface
$r=R$ between the metric~({\ref{tidalg}) and a general interior solution
of the form~(\ref{metric}) with $\lambda=\lambda(r)$ given by Eq.~\eqref{reglambda}
and $\nu=\nu(r)$ to be specified later.
This leads to the three equations
\be
\label{mgd00}
&&
e^{\nu_R}
=1-\frac{2\,\lp\,\cal{M}}{\mpl\,R}-\frac{q}{R^2}
\ ,
\\
\label{RegmatchNR1}
&&
\frac{2\,\cal{M}}{R}
=
\frac{2\,M}{R}-\frac{\mpl}{\lp}\left(f^{*}_R+\frac{q}{R^2}\right)
\ ,
\\
\label{RegmatchNR2}
&&
\frac{q}{R^4}
=
\left(\frac{\nu'_R}{R}+\frac{1}{R^2}\right)f^{*}_R+8\,\pi\,\frac{\lp}{\mpl}\,p_R
\ ,
\ee
where $\nu_R\equiv \nu_-(R)$, $\nu'_R\equiv \partial_r\nu_-|_{r=R}$,
$M=m(R)$ is the total GR mass of the star,
$p_R\equiv p(R)$ the pressure at the surface,
and $f^{*}_R=f^*(R;M,\sigma)$ encodes the MGD at $r=R$.
\par
In particular, the tidal charge $q$ in Eq.~\eqref{RegmatchNR2} can be
obtained from the second fundamental form given by Eq.~\eqref{matchingf},
which in our approach reduces to
\be
\frac{\lp}{\mpl}\,p_R
+\left(\frac{\nu'_R}{R}
+\frac{1}{R^2}\right)
f^*_R
= 
\frac{2\,{\cal U}_R^+}{k^4\,\sigma}
+\frac{4\,{\cal P}_R^+}{k^4\,\sigma}
\ .
\label{sffmgd}
\ee
On using the tidally charged metric~\eqref{tidalg} and the Weyl
functions~\eqref{exteriorWayl} in Eq.~\eqref{sffmgd},
we find the expression shown in Eq.~\eqref{RegmatchNR2}.
Now, from Eqs.~(\ref{RegmatchNR1}) and~(\ref{RegmatchNR2}),
we obtain the tidal charge as
\be
\frac{\mpl}{\lp}\,q
=
\left(\frac{R\,\nu'_R+1}{R\,\nu'_R+2}\right)
\left(\frac{2\,M}{R}-\frac{2\,{\cal M}}{R}\right)
R^2
+\frac{p_R\,R^4}{2+R\,\nu'_R}
\ ,
\label{qgeneral}
\ee
where $R\,\nu'_R$ in Eq.~\eqref{qgeneral} is given by setting
$r=R$ in the general BW expression
\be
r\,\nu'
=
\frac{k^2\,\tilde{p}\,r^2+2\,\tilde{m}/r}{1-2\,\tilde{m}/r}
\ ,
\label{BWexpression}
\ee
where $\tilde m$ was defined in Eq.~\eqref{massfunction} and the effective
BW pressure reads
\be
\tilde{p}
=
p+\frac{1}{\sigma}\left(\frac{\rho^2}{2}+\rho\, p
+\frac{2}{k^4}\,\cal{U}\right)
+\frac{4}{k^4}\frac{\cal{P}}{\sigma}
\ .
\label{BWpressure}
\ee
The above Eq.~\eqref{BWexpression} is obtained by
inserting the generic interior solution~\eqref{reglambda}
in the field equation~\eqref{ec2}.
\par
From Eqs.~\eqref{qgeneral}-\eqref{BWpressure},
it is clear that different interior solutions will produce different expressions
relating the tidal charge $q$ and ADM mass ${\cal M}$, which,
as we mentioned before, are usually treated as independent quantities.
With this information in hands, the next natural step is to consider
convenient interior solutions leading to a simple relationship
between $q$ and ${\cal M}$.
A simple enough relationship will also make it easier to study the
properties of the BH associated with the tidally charged
metric~\eqref{tidalg}, as we will show in due course.
\section{ADM mass and tidal charge}
\label{ADM_q}
Before we proceed, it is worth recalling a few properties we
expect $q$ should have.
First of all, if no source is present and ${\cal{M}}=0$, 
as well as in the GR limit $\sigma^{-1}\to 0$,
the tidal charge $q$ should vanish.
Secondly, $q$ should also vanish for very small star density,
that is for $R\to\infty$ at fixed $\cal M$ and $\sigma$.
\par
The simplest expression for $q$ holding the above fundamental
requirements is given by
\be
\label{qpeculiar}
q
=\frac{2\,K{\cal M}}{\sigma\,R}
\ ,
\ee
where $K$ is a constant with the same dimensions as $G_{\rm N}$.
The form~\eqref{qpeculiar} corresponds to the following value of 
the pressure at the star surface
\be
{4\,\pi\,R^3\,p_R}
=
\frac{\mpl\,{\cal M}\,K}{\lp\,\sigma\,R^2}
\left(2+R\,\nu'_R\right)
+({\cal M}-M)
\left(1+R\,\nu'_R\right)
\ ,
\label{bound2}
\ee
which, as it is well known, does not need to be zero in the BW.
To ensure an acceptable physical behaviour, in Eq.~\eqref{bound2}
we must have
\be
\left(\frac{M}{\cal M}-1\right)
<
\frac{\mpl\,K}{\lp\,\sigma\,R^2}\left(\frac{2+R\nu'_R}{1+R\nu'_r}\right)
\ .
\label{bound22}
\ee
Note that $p_R\to 0$ for $R\to\infty$ (at fixed $M$, $\mathcal M$
and $\sigma$), as well as for $\sigma^{-1}\to 0$ (at fixed $R$),
if ${\mathcal M}\to M$ in the same limit [in fact, see Eq.~(\ref{MMsM}) below].
In the simpler case $p_R=0$, the charge~\eqref{qpeculiar}
corresponds to the matching condition
\be
 R\,\nu'_R
=
-\frac{(M-{\cal M})-\frac{2\,{\cal M}\,K\,\mpl}{\sigma\,R^2\,\lp}}
{(M-{\cal M})-\frac{{\cal M}\,K\,\mpl}{\sigma\,R^2\,\lp}}
\ ,
\label{bound}
\ee
where we must have 
\be
\frac{{\cal M}\,K\,\mpl}{\sigma\,R^2\,\lp}
<
(M-{\cal M})
<
\frac{2\,{\cal M}\,K\,\mpl}{\sigma\,R^2\,\lp}
\label{cond}
\ee
to ensure an acceptable physical behaviour in the interior.
Thus, we may say that the matching conditions~\eqref{bound2} or \eqref{bound} 
lead to the simple exterior solution
\be
e^{\nu}
=1-\frac{2\,\lp\,{\cal M}}{\mpl\,r}-\frac{2\,K{\cal M}}{\sigma\,R\,r^2}
\ .
\label{simple}
\ee
The star radius $R$ in the solution (\ref{simple}) is still a free parameter.
However, it can be fixed for any specific star interiors,
as we will see in the next Section.
\par
First of all, notice from Eq.~\eqref{bound} that the ADM mass ${\cal M}$
can be written as 
\be
{\cal M}
=
\frac{M\,R^2}
{R^2+\frac{K\,\mpl}{\sigma\,\lp}\left(\frac{R\,\nu'_R+2}{R\,\nu'_R+1}\right)}
\ .
\label{Madmgeneral}
\ee
We may therefore consider different interior BW solutions yielding specific $R=R(M)$.
Secondly, in the MGD approach, we can always Taylor expand the geometric 
function $\nu=\nu(r;\sigma)$ in terms of the brane tension $\sigma$ as
\be
\nu(r;\sigma)
=
\nu_{\rm GR}(r)+\sigma^{-1}\,F^* + {\cal O}(\sigma^{-2})
\label{nusigma}
\ ,
\ee
where $\nu_{\rm GR}(r)$ is the GR solution of the geometric function $\nu$
and $F^*$ is a functional of the geometric deformation $f^*=f^*(r)$.
The above expansion implies that, at the star surface, $\nu'_R$ in
Eq.~\eqref{Madmgeneral} can be written as
\be
\nu'_R(\sigma)
=
\nu'_{\rm GR}(R)
+\left.\sigma^{-1}\,{F^*}'\right|_{r=R}
+{\cal O}(\sigma^{-2})
\label{nusigma2}
\ .
\ee
The ADM mass in Eq.~\eqref{Madmgeneral} can then be expressed as
\be
{\cal M}
=
\frac{M\,R^2}
{R^2+\frac{K\,\mpl}{\sigma\,\lp}\left[\frac{R\,\nu'_{\rm GR}(R)+2}{R\,\nu'_{\rm GR}(R)+1}\right]}
+{\cal O}(\sigma^{-2})
\ ,
\label{Madmgeneral2}
\ee
which shows that the GR solution $\nu_{\rm GR}$ for the interior geometric
function $\nu$ in Eq.~\eqref{Madmgeneral} determines corrections of order
$\sigma^{-1}$ in the BW ADM mass ${\cal M}$.
\par
In the following we shall always consider solutions with a linear
relationship,
\be
R
\propto
M
\ ,
\label{RpropM}
\ee
between the GR mass $M$ and star radius $R$.
\subsection{A linear BW star solution}
A BW solution with the property~\eqref{RpropM} was found in Ref.~\cite{jovalle207},
and is given by
\be
e^{\nu}=A\left(1+C\,r^2\right)^4
\ ,
\label{nubw1}
\ee
where the density is
\be
\rho(r)
=
C_\rho
\left(\frac{\mpl}{\lp}\right)
\frac{C\left(9+2\,C\,r^2+C^2\,r^4\right)}
{7\,\pi \,{\left( 1 + C\,r^2 \right) }^3}
\ ,
\label{Cdensity}
\ee
and the pressure is
\be
p(r)
=
\left(\frac{\mpl}{\lp}\right)
\frac{2\,C\left(2-7\,C\,r^2-C^2\,r^4\right)}{7\,\pi\left(1+C\,r^2\right)^3}
\ ,
\ee
with $A$ and $C_\rho$ are also constants.
Here $C$ is a function of the brane tension $\sigma$ and can always be expressed
as
\be
C(\sigma)=C_0+{\cal O}(\sigma)
\label{Csigma}
\ ,
\ee
where $C_0$ is the GR value of $C$, and is obtained by imposing $p_R=0$
(the GR value of the pressure at the star's surface).
\par
The ${\cal O}(\sigma)$ corrections are proportionals to the (minimal) geometric deformation
$f^*=f^*(r)$ and always depend on the matching conditions.
The fact that the function $C(\sigma)$ can be expanded as shown by Eq.~\eqref{Csigma}
implies that the geometric function $\nu$ can also be expanded as shown
in Eq.~\eqref{nusigma}, where the GR solution $\nu_{\rm GR}$ is obtained by using $C_0$
in Eq.~\eqref{nubw1}.
This solution yields
\be
R=
2\,n
\left(\frac{\lp}{\mpl}\right)
\frac{M}{C_\rho}
\ ,
\label{exac4}
\ee
with $n \equiv \frac{56}{43-\sqrt{57}}\simeq 1.6$. 
The general form of the ADM mass and tidal charge $q$ are then given by
\be
{\cal M}
=
\frac{M^3}
{M^2+\tilde n\,C_\rho^2\,K\,\frac{\mpl^2}{\mg^2}\,\frac{\mpl^3}{\lp}}
\ ,
\label{gMadm}
\ee
and
\be
q
=
\frac{C_\rho\,K\,\lg^2\,\mpl\,M^2}
{\tilde n_1\,\lp\,M^2
+\tilde n_2\,C_\rho^2\,K\,\frac{\mpl^2}{\mg^2}\,\mpl^3}
\ ,
\label{gqM}
\ee
where $\tilde n$, $\tilde n_1$ and $\tilde n_2$ are numerical coefficients
of order one, and $\sigma\simeq\lg^{-2}$ was used in the  expressions for
the ADM mass~\eqref{Madmgeneral} and tidal charge~\eqref{qpeculiar}.
\par
\begin{figure}[t]
\centering
\epsfxsize=8cm
\epsfbox{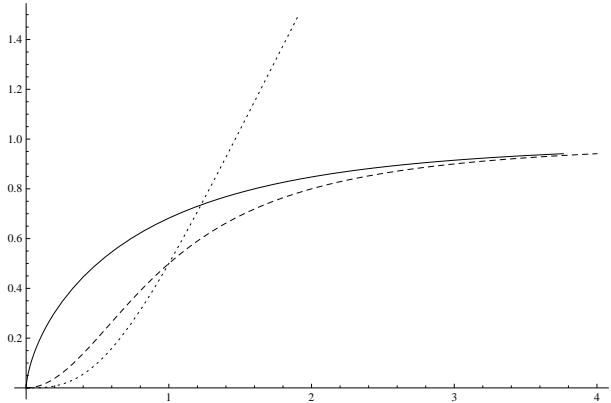}
\caption{Qualitative behaviour of the tidal charge $q=q({\cal M})$ (solid line),
$q=q(M)$ (dashed line), and ${\cal M}={\cal M}(M)$ (dotted line).
Units are arbitrary.
\label{qMM}}
\end{figure}
Note that Eq.~\eqref{gMadm} could now be used to express $M=M({\cal M})$
in Eq.~\eqref{gqM}, thus eliminating the (unmeasurable) $M$ from the
metric, and yielding $q=q({\cal M})$.
In particular, it is clear that the ADM mass equals $M$ for very large GR
mass $M$, 
\be
{\cal M}
\simeq
M
\ .
\label{MMsM}
\ee
The tidal charge also monotonically increases with $M$, but it asymptotes
to the constant
\be
q_\infty
=
\frac{C_\rho\,K\,\lg^2\,\mpl}
{\tilde n_1\,\lp}
\ .
\label{qinfty}
\ee
For the general qualitative behaviour of these quantities,
see Fig.~\ref{qMM}.
\par
Several choices of $K$ and $C_\rho$ can then be considered.
\subsubsection{The microscopic BH}
\label{microBH}
The choice  
\be
K
=
\frac{\lg}{\mg}
\left(\frac{\mpl}{\mg}\right)^2
\ ,
\qquad
C_\rho
&\!\!=\!\!&
\left(\frac{\mg}{\mpl}\right)^4
\ ,
\label{kc}
\ee
was studied in details in Ref.~\cite{covalle}, 
and leads to
\be
{\cal M}
=
\frac{M^3}
{M^2+n_1\,\mg^2}
\label{Madm}
\ee
and
\be
q
=
\frac{\lg^2\,M^2}
{n\,\left(M^2+n_1\,\mg^2\right)}
\ ,
\label{qM}
\ee
where ${n}_1 \simeq 0.14$.
Note that the asymptotic tidal charge~\eqref{qinfty} takes the rather
natural BW value
\be
q_\infty
=
\frac{\lg^2}{\tilde n_1}
=
\frac{\lg^2}{n\,n_1}
\simeq
\lg^2
\ ,
\label{qinftyLg}
\ee
and is therefore negligibly small for macroscopic stars.
This result means that a star of finite size will not probe the existence
of an extra-spatial dimension, unless its radius $R$ and mass $M$
are close to the typical length $\lg$ and mass $\mg$ of the RS model.
At this point the star density $\rho\sim\mg/\lg^3\sim\sigma$ and the tidal term
appears as a consequence of the truly five-dimensional nature of space-time.  
\par
The ADM mass~\eqref{Madm} and tidal charge~\eqref{qM} do not explicitly
depend on the star radius $R$, and we can therefore assume they remain
valid in the limit $R\to 0$ (or, more cautiously, $R\lesssim\lg$).
Correspondingly, the exterior metric given by Eq.~\eqref{simple} becomes
\be
e^{\nu}
=
1
-\frac{2\,\lp\,M^3}
{\mpl\left(M^2+n_1\,\mg^2\right) r}
\left(1
+\frac{\lg^2\,\mpl}
{2\,n\,\lp\,M\,r}
\right)
\ ,
\label{simple2}
\ee 
which can be used to describe a BH of ``bare'' (or GR) mass $M$.
This will be analysed in the last Section, along with other interior solutions.
\subsubsection{Gravitational screening}
\label{screening}
The choice  
\be
K
=
\frac{\lp}{\mpl}
\ ,
\qquad
C_\rho
=
1
\ ,
\label{kcs}
\ee
leads to the same asymptotic value~\eqref{qinftyLg} of the tidal charge,
which therefore implies that $q$ is practically negligible 
at all macroscopic scales.
Moreover,
\be
{\cal M}
=
\frac{M^3}
{M^2+\tilde n\,\frac{\mpl^2}{\mg^2}\,\mpl^2}
\ ,
\label{sMadm}
\ee
and
\be
q
=
\frac{\lg^2\,M^2}
{\tilde n_1\,M^2
+\tilde n_2\,\frac{\mpl^2}{\mg^2}\,\mpl^2}
\ ,
\label{sqM}
\ee
which shows that the ADM mass also becomes negligibly
small for $M\sim \mg$,
\be
{\cal M}
\sim
\left(\frac{\mg}{\mpl}\right)^4
M
\ll
M
\simeq
\mg
\ ,
\ee
and analogously for the tidal charge
\be
q
\sim
\lg^2\left(\frac{\mg}{\mpl}\right)^2
\left(\frac{M}{\mpl}\right)^2
\ll
\lg^2
\ .
\ee
\par
Since it is ${\cal M}$ which measures the gravitational strength of
a star in the weak field limit, one can view the above expressions as
describing a screening of the star's gravitational field for very small
GR mass.
In fact, this screening effect seems somewhat similar to the vanishing
of the ADM mass in the neutral shell model of Ref.~\cite{adm,cadm}.
\subsubsection{More general $K$ with $C_\rho=1$}
Keeping $C_\rho=1$, we shall now look at other expressions for $K$.
\par
For instance, the choice 
\be
K
=
\left(\frac{\lp}{\mpl}\right)\left(\frac{\mg}{\mpl}\right)^\alpha
\label{k7}
\ee
yields
\be
q_\infty
=
\frac{\lg^2}{\tilde n_1}\left(\frac{\mg}{\mpl}\right)^\alpha
\ ,
\ee
which can be very large or small depending on the sign of $\alpha$.
In particular, positive values of $\alpha$ imply a totally negligible
tidal charge, and we will consider this case just for completeness.
Further, 
\be
{\cal M}
=
\frac{M}
{1+\tilde n\,\left(\frac{\mg}{\mpl}\right)^{\alpha-2}\left(\frac{\mpl}{M}\right)^2}
\ ,
\ee
and
\be
q
=
\frac{(\mg/\mpl)^\alpha\,\lg^2}
{\tilde n_1+\tilde n_2\,\left(\frac{\mg}{\mpl}\right)^{\alpha-2}\left(\frac{\mpl}{M}\right)^2}
\ ,
\ee
which, for $M\sim \mg$, respectively become
\be
{\cal M}
\sim
\frac{M}
{1+\tilde n\,\left(\frac{\mg}{\mpl}\right)^{\alpha-4}}
\ ,
\ee
and
\be
q
\sim
\frac{(\mg/\mpl)^\alpha\,\lg^2}
{\tilde n_1+\tilde n_2\,\left(\frac{\mg}{\mpl}\right)^{\alpha-4}}
\ .
\ee
For $M\simeq \mg$, we then have
\be
{\cal M}
&\simeq&
\left(\frac{\mg}{\mpl}\right)^{4-\alpha}
\mg
\ll
\mg
\ ,
\qquad
\alpha<4
\ ,
\nonumber
\label{kp1}
\\
\\ \nonumber
{\cal M}
&\simeq&
\mg
\ ,
\qquad
\alpha\geq 4
\ ,
\nonumber
\ee
and
\be
q
&\simeq&
\lg^2\,\left(\frac{\mg}{\mpl}\right)^4
\ll
\lg^2
\ ,
\qquad
\alpha<4
\ ,
\nonumber
\label{kp2}
\\
\\ \nonumber
q
&\!\!\simeq\!\!&
\lg^2\,\left(\frac{\mg}{\mpl}\right)^\alpha
\ll
\lg^2
\ ,
\qquad
\alpha\geq 4
\ .
\nonumber
\ee
The first case ($\alpha<4$) therefore generalises the kind of screening effect
we described in Section~\ref{screening}, whereas the second case
($\alpha\geq 4$) would lead to a Schwarzschild-like exterior metric with ADM mass
of the order of $M\simeq \mg$.
However, as we noticed before, the tidal charge is always negligible for
$\alpha>0$ and the case $\alpha\ge 4$ is therefore of no physical interest,
since it practically coincides with the standard Schwarzschild geometry
for all GR masses $M\ge \mg$.
\par
Another choice we shall now describe is given by
\be
K=\left(\frac{\lg}{\mg}\right)\left(\frac{\mg}{\mpl}\right)^\beta
\ ,
\label{k77}
\ee
from which we obtain
\be
q_\infty
=
\frac{\lg^2}{\tilde n_1}\left(\frac{\mg}{\mpl}\right)^{\beta-2}
\ ,
\ee
which coincides with the previous case for $\beta=\alpha+2$.
In fact, the ADM mass 
\be
{\cal M}
=
\frac{M}
{1+\tilde n\,\left(\frac{\mg}{\mpl}\right)^{\beta-4}\left(\frac{\mpl}{M}\right)^2}
\ ,
\ee
and charge
\be
q
=
\frac{(\mg/\mpl)^{\beta-2}\,\lg^2}
{\tilde n_1+\tilde n_2\,\left(\frac{\mg}{\mpl}\right)^{\beta-4}\left(\frac{\mpl}{M}\right)^2}
\ ,
\ee
in the limit for $M\sim \mg$, yield
\be
{\cal M}
\sim
\frac{M}
{1+\tilde n\,\left(\frac{\mg}{\mpl}\right)^{\beta-6}}
\ ,
\ee
and
\be
q
\sim
\frac{(\mg/\mpl)^{\beta-2}\,\lg^2}
{\tilde n_1+\tilde n_2\,\left(\frac{\mg}{\mpl}\right)^{\beta-6}}
\ .
\ee
Again, in the limit $M\simeq \mg$, we obtain
\be
{\cal M}
&\simeq&
\left(\frac{\mg}{\mpl}\right)^{\beta-n}\,\mg
\ll
\mg
\ ,
\qquad
\beta<6
\ ,
\nonumber
\\
\label{kp3}
\\
\nonumber
{\cal M}
&\simeq&
\mg
\ ,
\qquad
\beta\geq 6
\nonumber
\ee
and
\be
q
&\simeq&
\lg^2\,\left(\frac{\mg}{\mpl}\right)^4
\ll
\lg^2
\ ,
\qquad
\beta<6
\ ,
\nonumber
\\
\\ \nonumber
q
&\!\!\simeq\!\!&
\lg^2\,\left(\frac{\mg}{\mpl}\right)^{\beta-2}
\ll
\lg^2
\ ,
\qquad
\beta\geq 6
\nonumber
\ ,
\ee
which reproduce the same behaviour as Eqs.~\eqref{kp1} and \eqref{kp2}.
The case of $\beta\ge 2$ again leads to negligible tidal charge and we
are just left with a large parameter range ($\beta<2$) with gravitational screening.
There is therefore hope that such a behaviour may represent a general feature
of (some) BW metrics in the limit $M\simeq \mg$.
\subsection{More linear interiors}
In order to assess the generality of the results obtained so far, we shall now
consider other interior BW solutions satisfying the condition~\eqref{RpropM},
or more specifically, solutions leading to Eq.~\eqref{exac4},
and therefore to Eqs.~\eqref{Madm}-\eqref{qM} when Eq.~\eqref{kc} is chosen. 
First of all, it is important to note that  the constant $n$ appearing in the
solution~\eqref{exac4} is proportional to the inverse of the compactness $M/R$.
Therefore any given value of the parameter $n$ for the interior BW
solution~\eqref{exac4} will represent a specific stellar distribution characterised
by a given compactness.
\par
We then consider two different BW interior solutions:
a non-uniform and a uniform stellar solution. 
\subsubsection{Non-uniform interior}
We will first consider the non-uniform stellar distribution found in Ref.~\cite{jovalle07},
which represents a physically acceptable solution to Eqs.~\eqref{ec1}-\eqref{con1}.
In fact, the corresponding metric is regular at the origin, with well-defined mass and radius, 
pressure and density are positive definite and decrease monotonically
with increasing radius, and so on.
The metric elements read
\be
\label{regularmet11}
e^{-\lambda(r)}=1-\frac{3\,C\,r^2}{2\left(1+C\,r^2\right)}+f^*(r)
\ ,
\ee
and
\be
\label{regularmet00}
e^{\nu(r)}=A(1+Cr^{2})^3
\ ,
\ee
corresponding to a density
\be
\label{regularden}
\rho(r)
=
\frac{3\,C\left(3+Cr^2\right)}{2\,k^2\left(1+Cr^2\right)^2}
\ ,
\ee
and pressure
\be
\label{regularpress}
p(r)
=
\frac{9\,C\left(1-Cr^2\right)}{2\,k^2\left(1+Cr^2\right)^2}
\ ,
\ee
with interior Weyl functions ${\cal U}$ and ${\cal P}$ given by
complicated expressions we shall not show here.
\par
The geometric deformation $f^*=f^*(r)$ in Eq.~\eqref{regularmet11}
is found by using Eqs.~\eqref{regularmet00}-\eqref{regularpress}
in Eq.~\eqref{fsolutionmin}) and $C=C(\sigma)$ as a function depending
on the geometric deformation evaluated at the star surface $r=R$.
The star radius is consequently given by Eq.~\eqref{exac4} with a specific
value for $n$, namely $n=4/3$ and $n_1\simeq 0.18$,
corresponding to a specific compactness of the stellar distribution.
When the relationship between $K$ and $C_\rho$ is given by Eq.~\eqref{kc}, 
this interior geometry leads to Eqs.~\eqref{Madm} and \eqref{qM},
so that the exterior geometry is again given by Eq.~\eqref{simple2}, 
which as before can be used to describe a BH of ``bare'' mass $M$
(see last Section).
\subsubsection{The Schwarzschild BW solution}
In GR, the well-known Schwarzschild interior metric,
representing a uniform stellar distribution, is the only stable solution
for a bounded distribution~\cite{weinberg} which matches smoothly
with the Schwarzschild exterior metric.
It is also important to note that, in the BW, the collapse of
a uniform star leads to a non-static exterior solution~\cite{BGM,dad,kp2004}.
Nonetheless, the study of this interior solution in the BW context 
represents a scenario of great interest.
Indeed, in the pioneering work of Germani and Maartens~\cite{gm},
a BW solution for the Schwarzschild's interior solution was reported,
but only high-energy corrections were considered, 
leaving out the effects of Weyl functions inside uniform distributions,
which we now know are necessary for a consistent description. 
\par
When the Schwarzschild interior solution is considered in 
the MGD approach, using local high-energy corrections along
with non-local bulk terms, a consistent BW solutions can indeed be found,
showing that non-local effects play a relevant role inside the stellar distribution.
Indeed, both Weyl interior functions ${\cal U}$ and ${\cal P}$ are in general
proportional to the compactness of the stellar distribution.
This metric reads
\be
\label{schw11}
e^{-\lambda}=1 - \frac{r^2}{C^2}+f^*(r)
\ ,
\ee
and
\be
\label{schw00}
e^{\nu}=\left(A-B\sqrt{1-\frac{r^2}{C^2}}\right)^2
\ ,
\ee
with density
\be
\label{schwdensity}
\rho =\frac{3}{k^2\, C^2 }
\ ,
\ee
and pressure
\be
\label{schwpressure}
p(r)
=
\frac{\rho}{3}
\left[
\frac{{3\,B\,\sqrt{1 -\frac{r^2}{C^2}}} - A}
{A-B\,\sqrt{1 -\frac{r^2}{C^2}}}
\right]
\ .
\ee
The interior Weyl functions ${\cal U}$ and ${\cal P}$ are again omitted
for the sake of simplicity, and the function $f^*=f^*(r)$ in Eq.~\eqref{schw11}
is the geometric deformation found by using
Eqs.~\eqref{schw00}-\eqref{schwpressure} in Eq.~\eqref{fsolutionmin}.
\par
This solution represents a BW uniform compact stellar distribution
and leads to Eq.~\eqref{exac4} with a value for $n$ depending
on the matching conditions at $r=R$.
For instance, for $p_R=0$, $A=1$ and $B=1/2$, we obtain $C^2=(9/5)\,R^2$,
leading to $n=9/5$ and $n_1\simeq 0.1$, again corresponding to a specific
compactness.
When the relationship between $K$ and $C_\rho$ is given by Eq.~\eqref{kc}, 
this interior geometry leads to  Eqs.~\eqref{Madm} and \eqref{qM},
so that the exterior geometry is again given by Eq.~\eqref{simple2}, 
which as before can be used to describe a BH of ``bare'' mass $M$. 
\section{BH limit and minimum mass}
\label{bh}
We shall now review the BH limit for the case in Section~\ref{microBH},
as previously analysed in Ref.~\cite{covalle}, and compare with what can be obtained
from the other star solutions described above.
\par 
Introducing the dimensionless proper mass $\bar M={M}/{\mg}$ in
Eqs.~\eqref{Madm} and \eqref{qM}, we obtain
\be
\bar{\cal M}
=
\frac{\cal M}{\mg}
=
\frac{\bar M^3}
{\bar M^2+{n}_1}
\simeq
\frac{\bar M^3}
{0.14+\bar M^2}
\ ,
\label{bMM}
\ee
and the corresponding dimensionless tidal charge
\be
\bar q
=
\frac{q}{\lg^2}
=
\frac{\bar M^2}
{n\,({n}_1+\bar M^2)}
\simeq
\frac{\bar M^2}{0.22+1.6\,\bar M^2}
\ .
\label{bq}
\ee
From Eq.~\eqref{qM}, an important result for microscopic BHs was inferred.
The condition that is usually taken to define a semiclassical BH is that its horizon
radius $\rh$ be larger than the Compton wavelength
$\lambda_M\simeq \lp\,\mpl/M$ (see~\cite{acmo} and References therein),
\be
\rh\gtrsim\lambda_M
\ .
\label{semiBH}
\ee
From Eq.~(\ref{tidalg}), we obtain the horizon radius
\be
\rh
=
\frac{\lp}{\mpl}\left({\cal M}+\sqrt{{\cal M}^2+q\,\frac{\mpl^2}{\lp^2}}\right)
\ ,
\ee
and the classicality condition~\eqref{semiBH} reads
\be
\frac{M}{\mpl^2}
\left({\cal M}+\sqrt{{\cal M}^2+q\,\frac{\mpl^2}{\lp^2}}\right)
\gtrsim
1
\ .
\ee
We define the critical mass $\mc$ as the value of $M$
which saturates the above bound.
In order to proceed, we shall expand for $M\sim{\cal M}\simeq \mg\ll\mpl$,
thus obtaining
\be
\frac{\rh^2}{\lambda_{M}^2}
\simeq
\frac{M^2}{\mpl^2}\frac{q}{\lp^2}
\simeq
\frac{\mg^2}{\mpl^2}\,\bar M^2\,\bar q\,\frac{\lg^2}{\lp^2}
\simeq
\bar M^2\,\bar q
\simeq
1
\ ,
\ee
or
\be
{\bar M}^4-n\,{\bar M}^2-n\,n_1\simeq 0
\ ,
\label{eqMc}
\ee
which yields
\be
\mc\simeq 1.3\,\mg
\ ,
\label{Mc1}
\ee
or ${\mathcal M}_{\rm C}\simeq 1.2\,\mg$, from Eq.~\eqref{bMM}.
This can be viewed as the minimum allowed mass for a semiclassical
BH in the BW~\cite{covalle}.
\par
When the non-uniform BW solution given by
Eqs.~\eqref{regularmet11})-\eqref{regularpress} is considered,
the corresponding values $n=4/3$ and $n_1\simeq 0.18$ used in
Eqs.~\eqref{bMM} and~\eqref{bq} yield
\be
\bar{\cal M}
\simeq
\frac{\bar M^3}
{0.18+\bar M^2}
\label{bMM2}
\ee
and
\be
\bar q
\simeq
\frac{\bar M^2}{0.24+1.3\,\bar M^2}
\ .
\label{bq2}
\ee
Therefore, this non-uniform BW solution can also be associated with a minimum
mass, allowing for the existence of semiclassical BHs in the BW, given by
\be
\mc\simeq 1.2\,\mg
\ .
\label{Mc2}
\ee
The compactness is the main difference between the two non-uniform stellar
distributions studied here, and this seems the most likely cause of the slight
difference between the critical mass shown in Eq.~\eqref{Mc1}
(less compact) and Eq.~\eqref{Mc2} (more compact).
\par
On the other hand, Eqs.~\eqref{bMM} and \eqref{bq} evaluated for
the Schwarzschild BW solution given in Eqs.~\eqref{schw11}-\eqref{schwpressure},
with $n=9/5$ and $n_1\simeq 0.1$, yield
\be
\bar{\cal M}
\simeq
\frac{\bar M^3}
{0.1+\bar M^2}
\label{bMM3}
\ee
and
\be
\bar q
\simeq
\frac{\bar M^2}{0.18+1.8\,\bar M^2}
\ .
\label{bq3}
\ee
Hence, the corresponding minimum mass for a semiclassical 
BH in the BW would now be given by the somewhat larger 
\be
\mc\simeq 1.9\,\mg
\ .
\ee
We note that the Schwarzschild solution has the largest value for $n$,
that is, it represents the less compact distribution among the three solutions
discussed here.
Again, we found evidence showing that less compact distributions correspond
to greater critical mass $\mc$. 
\par
We can conclude that different values of the compactness for the
star we assume as the starting point of the mathematical limit leading
to a BH result in different minimum allowed masses for semiclassical BHs.
It is already clear that the compactness of the stellar solution
associated with the interior geometry plays a fundamental role
in the study of the BH showing up from the exterior solution.
Indeed, when solutions satisfying the condition~\eqref{exac4} are considered, 
therefore leading to Eqs.~\eqref{Madm}-\eqref{qM} when Eq.~\eqref{kc} is chosen, 
we can find a general result relating the compactness of the interior solution
and the minimum allowed mass.
From saturating the condition~\eqref{semiBH}, that is $\rh\simeq\lambda_M$,
we find ${\bar M}^2{\bar q}\simeq 1$, which yields Eq.~\eqref{eqMc}. 
Hence,
\be
{\bar M}^2_{\rm C}
\simeq
\frac{n}{2}\left(1+\sqrt{1+\frac{4\,n_1}{n}}\right)
\ge
n
\ .
\ee
Now, from GR we know that the compactness of any stable stellar distribution
of mass $M$ and radius $R$ must satisfy the constraint $M/R<4/9$. 
This bound in Eq.~\eqref{exac4} leads to $n>9/8$, and
\be
{\bar M}^2_{\rm C}
>
1
\ .
\ee
We must therefore have $\mc>\mg$,
and the linear relation~\eqref{exac4} between $M$ and $R$
will always lead to a critical mass $\mc$ above $\mg$.
\section{Conclusions}
\label{conc}
In this paper we have constructed analytical models of spherically symmetric stars
in the BW, such that the external space-time contains both an ADM mass ${\cal M}$ and
a tidal charge $q$.
In order to determine the exterior geometry, namely, to satisfy the physical
requirements for a general relationship among the ADM mass ${\cal M}$,
tidal charge $q$ and brane tension $\sigma$, we employed Israel's matching
conditions at the star surface $r=R$ on the brane, and considered several
candidate BW interior solutions with total GR mass $M$.
In particular, all the interior solutions were obtained from the principle of MGD
applied to known GR solutions. 
\par
Our subsequent analysis was restricted to stars with a radius $R$ linearly related to the
total GR mass $M$, thus resulting in a relatively simple relation between $M$,
the BW ADM mass ${\cal M}$, the tidal charge $q$,
and two (arbitrary) dimensional constants, $K$ and $C_\rho$.
These constants were conveniently chosen so as to generate 
manageable expressions and (hopefully) physically relevant solutions.
An interesting feature emerged then that, in all of the models we have analysed,
the value of the star's radius can be taken to zero smoothly, without
encountering singular expressions for ${\cal M}={\cal M}(M)$ or $q=q(M)$.
This mathematical procedure therefore yields BW BH metrics with a tidal
charge solely determined by the mass of the source $M$, or equivalently
${\cal M}$ (and the brane tension).
\par
Of course, the details of our models depend on the choice of those two
constants $K$ and $C_\rho$.
The first one [defined in Eq.~\eqref{qpeculiar}] is simply a coefficient of
proportionality between $q$ and ${\cal M}$,
whereas the second one [introduced in Eq.~\eqref{Cdensity}]
should be related to the likely typical density of a star distribution for
the specific problem one is addressing.
For regular stars, this parameter can therefore be fixed by comparing
with available astrophysical data.
Alternatively, for the purpose of describing BHs in the BW, $C_\rho$ should be
chosen so as to model the properties of BW matter in the limit of
vanishing star radius, $R\ll\lg$ (or, more realistically, at the end-point of
gravitational collapse).
Together, $K$ and $C_\rho$ determine the asymptotic value $q_\infty$ of the
tidal charge for stars of very large GR mass $M$ (or, equivalently, ADM mass ${\cal M}$).
The simplest choice, namely $K=\lp/\mpl$ and $C_\rho=1$ (for which $q_\infty\simeq\lg^2$), 
was first shown in Section~\ref{screening} to yield an interesting gravitational
screening effect, namely ${\cal M}\ll\mg$ for $M\simeq \mg$, which resembles
the vanishing of the ADM mass in the GR neutral shell model of Ref.~\cite{adm,cadm}.
In fact, even in this model of fundamental particles as worm-holes, the proper
mass $M$ remains finite in the ``point-like'' limit, but the corresponding ADM
mass vanishes~\footnote{More generally, ${\cal M}$ is proportional to the
particle's electric charge~\cite{adm}.
Charged BW stars will therefore be considered in a future publication.}.
Keeping $C_\rho=1$, more general forms of $K$ were also considered,
always yielding a tidal charge $q\ll\lg^2$ for $M \simeq \mg$.
Further, the gravitational screening was seen to occur again in a whole
family of cases [see Eqs.~\eqref{kp1} with $\alpha<0$ and \eqref{kp3}
with $\beta<2$].
\par
The specific choice of $K$ and $C_\rho$ in Section~\ref{microBH}
represents the microscopic BH case previously studied in Ref.~\cite{covalle},
where a minimum (critical) mass $\mc$ for semiclassical BHs was derived.
Two more BW solutions were then considered in this limit, a non-uniform one
and the uniform Schwarzschild BW interior, corresponding to different
star compactness $M/R$.
It was then found that different ratios $M/R$ produce different minimum allowed
mass $\mc$ for semiclassical BHs, although the magnitude of the variation is not
particularly significant, since one always finds $\mc\gtrsim\mg$.
In fact, we were able to show that, assuming the simple linear relation between
$M$ and $R$ given in Eq.~\eqref{exac4}, and the well-known GR constraint
$M/R<4/9$, always lead to a critical mass $\mc$ above $\mg$.
This conclusion may have a significant impact on the search for microscopic
BHs in the BW~\cite{CH,cavaglia}.
\par
We wish to conclude by emphasising that our analysis has a natural
application to the study of BW stars and that the models we built here
are mainly intended for astrophysical applications.
In this respect, the main result is therefore that the tidal charge~\eqref{gqM}
is almost certainly negligible for regular stars, and the simple
Schwarzschild metric can be effectively used for the exterior of
stars in the BW.
One may further view the present findings as a first step toward
a realistic description of such objects.
The ambiguities associated with the various approximations employed
could next be resolved by confronting with phenomenological data.
%
%
%
%
%
%
%
%
%
%
%

%
\end{document}